\documentstyle[tighten,preprint,eqsecnum,aps]{revtex}
\begin{document}
\draft

\hyphenation{
mani-fold
mani-folds
}


\def\half{{1\over2}}
\def\casehalf{{\case{1}{2}}}
\def\onethird{{1\over3}}
\def\caseonethird{{\case{1}{3}}}

\def\mink{{M^{2+1}}}

\def\bbase{{\cal B}}

\def\so{{\hbox{${\rm SO}(2,1)$}}}
\def\soc{{\hbox{${\rm SO}_0(2,1)$}}}
\def\su{{\hbox{${\rm SU}(1,1)$}}}

\def\tsoc{{\hbox{${\widetilde{\rm SO}_0}(2,1)$}}}

\def\iso{{\hbox{${\rm ISO}(2,1)$}}}
\def\isoc{{\hbox{${\rm ISO}_0(2,1)$}}}
\def\isu{{\hbox{${\rm ISU}(1,1)$}}}

\def\sltwoz{{\hbox{${\rm SL}(2,{\bf Z})$}}}
\def\sltwor{{\hbox{${\rm SL}(2,{\bf R})$}}}

\def\miso{{{\cal M}}}
\def\ms{{{\cal M}_s}}
\def\mt{{{\cal M}_t}}
\def\mn{{{\cal M}_n}}
\def\mo{{{\cal M}_0}}

\def\bmt{{{\overline{\cal M}}_t}}
\def\bms{{{\overline{\cal M}}_s}}

\def\nisu{{{\cal N}}}
\def\ns{{{\cal N}_s^{\eta_1\eta_2}}}
\def\nt{{{\cal N}_t}}
\def\nn{{{\cal N}_n^{\eta_1\eta_2}}}
\def\no{{{\cal N}_0^{\eta_1\eta_2}}}

\def\bnt{{{\overline{\cal N}}_t}}

\def\teich{Teichm\"uller}


\def\mathpounds{{\mathchoice{{\hbox{\pounds}}}{{\hbox{\pounds}}}%
{{\hbox{\scriptsize\pounds}}}{{\hbox{\scriptsize\pounds}}}}}


\preprint{\vbox{\baselineskip=12pt
\rightline{SU-GP-93/7-6}
\rightline{CGPG-93/8-3}
\rightline{gr-qc/9308018}}}
\title{Solution space of 2+1 gravity on ${\bf R} \times T^2$ \\
in Witten's connection formulation}
\author{Jorma Louko\cite{jorma} and Donald M. Marolf\cite{don}}
\address{
Department of Physics, Syracuse University,
Syracuse, New York 13244--1130, USA
}
\date{Revised version, October 1993}
\maketitle
\begin{abstract}
We investigate the space $\miso$ of classical solutions to Witten's
formulation of 2+1 gravity on the manifold ${\bf R} \times T^2$. $\miso$
is connected, unlike the spaces of classical solutions in  the cases where
$T^2$ is replaced by a higher genus  surface. Although $\miso$ is neither
Hausdorff nor a manifold, removing from $\miso$ a set of measure zero
yields a manifold which is naturally viewed as the cotangent bundle over a
non-Hausdorff base space~$\bbase$. We discuss the relation of the various
parts of $\miso$ to spacetime metrics, and  various possibilities of
quantizing~$\miso$. There exist quantizations in which the exponentials of
certain momentum operators, when operating on states whose support is
entirely on the part of $\bbase$ corresponding to conventional spacetime
metrics, give states whose support is entirely outside this part
of~$\bbase$. Similar results hold when the gauge group $\soc$ is replaced
by $\su$.

\end{abstract}

\pacs{Pacs: 04.60.+n, 04.20.Jb}

\narrowtext

\section{Introduction}
\label{sec:intro}

The observation that vacuum Einstein gravity in 2+1 spacetime dimensions
has no local dynamical degrees of freedom\cite{deser1,deser2} has created
interest in 2+1 gravity as an arena where quantum gravity can be
investigated without many of the technical complications that are
present in 3+1 spacetime dimensions. Of particular interest for the 3+1
theory is to understand the relation between the various 2+1 quantum
theories that have been constructed in the metric, connection, and loop
formulations
\cite{martinec,witten1,bengtsson1,five-a,AAbook2,hosoya-nakao2,%
carlip1,carlip2,carlip3,mano,barbero}.
For a recent review, see Ref.\cite{carlip-water}.

In this paper we shall be interested in Witten's connection
formulation\cite{witten1} of 2+1 gravity and its relation to spacetime
metrics on spacetimes of the form ${\bf R}\times \Sigma$ where $\Sigma$
is a closed orientable surface of genus~$g$. The case $g>1$ is well
understood: the space of classical solutions to Witten's theory contains
several disconnected components, one of which is the cotangent bundle over
the \teich\ space of~$\Sigma$, and this component is isomorphic to the
solution space of the conventional metric
formulation\cite{witten1,goldman1,goldman2,moncrief,mess}. The case $g=0$,
corresponding to $\Sigma=S^2$, is also well understood: it is trivial, in
the sense that Witten's theory possesses only one classical solution and
the conventional metric formulation possesses no
solutions\cite{moncrief,mess}. For the remaining case $g=1$, corresponding
to $\Sigma=T^2$, the conventional metric formulation is well
understood\cite{moncrief,mess,hosoya-nakao1} and partial discussions of
Witten's formulation have been given in
Refs.\cite{five-a,AAbook2,carlip1,marolf}. However, a complete analysis of
the space $\miso$ of classical solutions to Witten's formulation for $g=1$
appears never to have been given. The purpose of this paper is to provide
such an analysis, and to explore the resulting possibilities for quantizing
the theory.

One motivation for looking at this question, besides its intrinsic
interest, is that the case $g=1$ has been much invoked as a testing ground
for calculations which for $g>1$ would require considerably more
sophistication. An example is the analysis of the relation of Witten's
connection quantization to metric quantization in
Refs.\cite{carlip1,carlip2,carlip3,anderson}. Another example is the
analysis of the Rovelli-Smolin loop transform in
Refs.\cite{five-a,AAbook2,marolf}. It is clearly important to understand
to what extent such analyses represent features that are present also in
the case $g>1$.

Conventional metric formulation of 2+1 gravity on ${\bf R}\times
T^2$ assumes that the three-metric is nondegenerate and that there exists a
foliation in which the induced metrics on the tori are
spacelike. Such spacetimes are known to correspond to roughly speaking half of
the space $\miso$ of classical solutions in Witten's
formulation\cite{moncrief,mess}. We shall however see that all points
in~$\miso$, except a set of measure zero, do correspond to nondegenerate
spacetime metrics on ${\bf R}\times T^2$. The nondegenerate spacetimes arising
from those points of $\miso$ that do not correspond to the conventional metric
formulation possess no slicing in which the induced metrics on the tori would
be spacelike.

The main difference
between $g=1$ and the higher genus case is that $\miso$ is {\it
connected,} even though it is neither Hausdorff nor a manifold. This
raises various possibilities for quantizing~$\miso$.
We shall see that removing from $\miso$ a suitable subset of measure zero
yields a manifold which can be naturally viewed as the cotangent bundle
over a non-Hausdorff base space~$\bbase$. Quantization of this cotangent
bundle yields a quantum theory which contains all the individual quantum
theories previously presented in Refs.\cite{five-a,AAbook2,carlip1};
these smaller theories each correspond to quantizing roughly speaking
half of~$\miso$. Further, the larger quantum theory contains operators
which induce transitions between the theories of
Refs.\cite{five-a,AAbook2,carlip1}. Such operators can in particular take
states whose support is entirely on the part of $\bbase$ which corresponds
to conventional spacetime metrics
and map them to states whose support is
entirely outside this part of~$\bbase$.

On a par with Witten's theory, we shall consider its modification in which
the gauge group $\soc$ (the connected component of $\so$) is replaced by
its double cover $\su\simeq\sltwor$. Although the technical details differ,
the overall qualitative picture is highly similar, and the space of
classical solutions in the modified theory will be just a four-fold cover
of~$\miso$. One reason for studying the modified theory is that this
theory is closer to the work done with the 3+1 loop transform in Ashtekar's
variables\cite{rov-smolin,AAbook-loop}. Also, the work done with the 2+1
loop transform in Refs.\cite{five-a,AAbook2} is perhaps more appropriately
understood in terms of the modified theory, and  the detailed study of the
2+1 loop transform in Ref.\cite{marolf} was explicitly carried out within
the modified theory.

The rest of the paper is as follows. In section \ref{sec:action} we
briefly outline Witten's formulation (and its $\su$ modification) of
2+1 gravity on a spacetime with topology ${\bf R}\times\Sigma$. In
section \ref{sec:solutions} we specialize to the case $\Sigma=T^2$ and
give a detailed analysis of the space of classical solutions. The relation
of the connection formulation to spacetime metrics is examined in
section~\ref{sec:metrics}, and various avenues for quantization are
explored in section~\ref{sec:quantization}. Section \ref{sec:discussion}
contains a brief summary and discussion.

\section{Connection dynamics}
\label{sec:action}

We consider a formulation of 2+1 gravity \cite{witten1,AAbook2,romano} in
which the fundamental variables of the theory are the co-triad ${\bar
e}_{aI}$ and the connection ${\bar A}_a^I$ defined on a three-dimensional
manifold~$M$.  The connection takes values in the Lie algebra of $\so$,
and the co-triad takes values in the dual of this Lie algebra. The internal
indices $I,J,\ldots$ thus take values in $\left\{0,1,2\right\}$, and they
are raised and lowered by the internal Minkowski metric,
$\eta_{IJ}={\rm{diag}}(-1,1,1)$. The abstract indices $a,b,\ldots$ are
spacetime indices on~$M$. We shall regard ${\bar A}_a^I$ either as an
$\so$ connection or as an $\su$ connection. The differences between
the two cases will become clear in the later sections.

The dynamics of the system is derived from the action
\begin{equation}
S \left( {\bar e}_{aI} , {\bar A}_a^I \right)
= \casehalf \int_{M} {d^3x} \,
\tilde{\eta}^{abc}
\, {\bar e}_{aI} {\bar F}^I_{bc}
\ \ ,
\label{3-action}
\end{equation}
where $\tilde{\eta}^{abc}$ is the Levi-Civita density on~$M$ and
${\bar F}^I_{bc}$ is the curvature of the connection,
\begin{equation}
{\bar F}^I_{bc} = 2 \partial^{\phantom{I}}_{[b} {\bar A}^I_{c]} +
\epsilon^I{}_{JK} {\bar A}^J_b {\bar A}^K_c
\ \ .
\end{equation}
The structure constants $\epsilon^I{}_{JK}$ are obtained from the totally
antisymmetric symbol $\epsilon_{IJK}$ by raising the index with the
Minkowski metric. Our convention is $\epsilon_{012}=1$. For later
convenience, we have followed Ref.\cite{romano} and taken the action
(\ref{3-action}) to differ from that adopted in
Refs.\cite{witten1,carlip1,carlip2,carlip3} by a factor of~$\casehalf$.
The action may need to be supplemented with boundary terms depending on
what is held fixed in the variational principle. However, we shall not
need to discuss these boundary terms here.

The classical equations of motion are
\begin{mathletters}
\label{eom3d}
\begin{eqnarray}
&&{\bar F}^I_{ab} = 0
\\
&&{\bar {\cal D}}_{[a} {\bar e}_{b]I} = 0
\ \ ,
\label{3-eom2}
\end{eqnarray}
\end{mathletters}
where ${\bar {\cal D}}_a$ is the gauge covariant derivative determined by
${\bar A}^I_a$,
\begin{equation}
{\bar {\cal D}}_a v_K =
\partial_a v_K - \epsilon^I{}_{JK} {\bar A}^J_a v_I
\ \ .
\label{dbar}
\end{equation}
Note that ${\bar {\cal D}}_a$ acts only on the internal indices.
Equations (\ref{eom3d}) therefore say that the connection ${\bar
A}^I_a$ is flat and the co-triad is compatible with the connection.
If the co-triad is nondegenerate,  the metric
$g_{ab}= {\bar e}_{aI} {\bar e}_{b}^I$ has signature $(-,+,+)$, and the
equations of motion imply that this metric is flat, {\it i.e.}, satisfies the
vacuum Einstein equations. However, the theory remains well-defined also for
degenerate co-triads.

We now take $M$ to be ${\bf R}\times\Sigma$, where $\Sigma$ is
a closed orientable two-manifold. The 2+1 decomposition of the action takes
the form\cite{witten1,AAbook2,romano}
\begin{equation}
S = \int dt \int_\Sigma d^2x
\left[
{\tilde E}^j_I \left( \partial_t A^I_j \right) +
{\bar A}^I_t \left( {\cal D}_j {\tilde E}^j_I \right)
+ \casehalf {\bar e}_{tI} F^I_{ij} {\tilde \eta}^{ij}
\right]
\ \ .
\label{2+1-action}
\end{equation}
The abstract indices $i,j,\ldots$ live on~$\Sigma$, and $t$ is the
coordinate on~${\bf R}$. The $\so$ or $\su$ connection $A^I_j$ is
the pull-back of ${\bar A}_a^I$ to~$\Sigma$, $F^I_{ij}$ is its curvature
given by
\begin{equation}
F^I_{ij} = 2 \partial^{\phantom{I}}_{[i} A^I_{j]} +
\epsilon^I{}_{JK} A^J_i A^K_j
\ \ ,
\end{equation}
and ${\tilde \eta}^{ij}$ is the Levi-Civita density on~$\Sigma$.
The vector density ${\tilde E}^j_I$ is given by ${\tilde E}^j_I={\tilde
\eta}^{ji} e_{iI}$, where $e_{iI}$ is the pull-back of ${\bar e}_{aI}$
to~$\Sigma$. ${\cal D}_j$ is the gauge-covariant derivative on $\Sigma$
determined by~$A^I_j$,
\begin{equation}
{\cal D}_j v_K =
\partial_j v_K - \epsilon^I{}_{JK} A^J_j v_I
\ \ .
\end{equation}
Note that the decomposition involves no assumptions about the rank of the
metric $g_{ab}= {\bar e}_{aI} {\bar e}_{b}^I$, or about the sign of the
quantity
${\bar e}_{tI} {\bar e}_{t}^I = g_{ab} {(\partial/\partial t)}^a
{(\partial/\partial t)}^b$\cite{romano}. This means that the induced metric on
$\Sigma$, $e_{iI}e_j^I$, can (locally) have any of the signatures $(+,+)$,
$(-,+)$, $(0,+)$, $(-,0)$, or $(0,0)$.

The canonical pair is now $\left(A^I_j,{\tilde E}^{j}_{I}\right)$,
with the Poisson brackets
\begin{equation}
\left\{A^I_i(x) , {\tilde E}^j_J (x')
\right\} = \delta^j_i \delta^I_J \delta^2 \left(x,x'\right)
\ \ ,
\label{AE-PB}
\end{equation}
where $x$ denotes the points on~$\Sigma$. ${\bar e}_{tI}$ and ${\bar
A}^I_t$ act as Lagrange multipliers enforcing the constraints
\begin{mathletters}
\label{constraints}
\begin{eqnarray}
F^I_{ij} &=& 0
\label{F-constraint}
\\
{\cal D}_j {\tilde E}^j_I &=& 0
\ \ .
\label{G-constraint}
\end{eqnarray}
\end{mathletters}
The first of these constraints generates analogues of the Hamiltonian
gauge transformations of the metric formulation of 2+1 gravity and their
extensions to degenerate co-triads. The second constraint is analogous to
the Gauss law constraint in electrodynamics. It generates internal
$\soc$ or $\su$ gauge transformations, depending on whether
${\bar A}_a^I$ is an $\so$ connection or an $\su$ connection.

When ${\bar A}_a^I$ is an $\so$ connection, the pair
$\left(A^I_j,e_{Ij}\right)$ defines on $\Sigma$ the $\iso$ connection
$e^I_j P_I + A^I_j K_I$,
where $P_I$ and $K_I$ are the generators of translations and Lorentz
transformations in 2+1 dimensional Minkowski space, satisfying the
commutator algebra
\begin{equation}
\begin{array}{rcl}
\left[ K_I , K_J \right] &=& \epsilon^K{}_{IJ} K_K
\\
\left[ K_I , P_J \right] &=& \epsilon^K{}_{IJ} P_K
\\
\left[ P_I , P_J \right] &=& 0
\ \ .
\end{array}
\end{equation}
Witten \cite{witten1} observed that the constraints (\ref{constraints})
make the curvature of this $\iso$ connection vanish. Further, these
constraints generate local $\isoc$ gauge transformations, where
$\isoc$ is the connected component of $\iso$.
Taking the connection to be in the trivial principal bundle
$\isoc\times\Sigma$, this means that the space of classical solutions is the
moduli space of flat $\isoc$ connections on~$\Sigma$, modulo $\isoc$ gauge
transformations.
When ${\bar A}_a^I$ is an $\su$ connection the situation is analogous, and
the space of classical solutions is the moduli space of flat $\isu$
connections on~$\Sigma$, modulo $\isu$ gauge transformations.

A flat connection on $\Sigma$ is completely determined by its holonomies
around closed noncontractible loops. The space of classical solutions is
therefore the space of group homomorphisms from the fundamental group of
$\Sigma$ to~$G$, where $G$ is respectively $\isoc$ or $\isu$,
modulo conjugation by~$G$. In the case $\Sigma=S^2$ the fundamental group
is trivial, and the space of classical solutions consists of only one
point. In all cases where $\Sigma$ is a surface of genus two or higher
the spaces of classical solutions are nontrivial and can be described
in a unified manner\cite{witten1,goldman1,goldman2}. In the remaining case,
$\Sigma=T^2$, the space of classical solutions is nontrivial but
qualitatively different from those of the higher genus surfaces. Our aim
is to analyze this remaining case in detail.

\section{Space of classical solutions for $\Sigma=T^2$}
\label{sec:solutions}

In this section we shall describe the space of classical solutions for
$\Sigma=T^2$. As the fundamental group of the torus is the Abelian group
${\bf Z}\times {\bf Z}$, we see from the end of the previous section that
points in the space of classical solutions are just equivalence classes of
pairs of commuting elements of~$G$ under $G$ conjugation. We shall devote
separate subsections to the cases $G=\isoc$ and $G=\isu$.

\subsection{$G=\isoc$}
\label{subsec:so21}

We first consider the case where ${\bar A}_a^I$ is an $\so$ connection.
Then $G$ is $\isoc$, that is, the group of proper Poincare
transformations in 2+1 dimensional Minkowski space~$\mink$. We denote the
space of classical solutions by~$\miso$.

Recall that $\isoc$ can be defined as the group of pairs $(R,w)$, where
$R$ is an $\soc$ matrix and $w$ is a column vector. The group
multiplication law is
\begin{equation}
(R_2,w_2) \cdot (R_1,w_1) =
(R_2 R_1, R_2 w_1 + w_2)
\ \ .
\end{equation}
When points in $\mink$ are represented by column vectors as
$v={(T,X,Y)}^T$ and the entries are the usual Minkowski coordinates
associated with the line element $ds^2 = - dT^2 + dX^2 + dY^2$, the action
of a group element $(R,w)$ on $\mink$ is
\begin{equation}
(R,w): \ \ v \mapsto Rv + w
\ \ .
\end{equation}

We now wish to obtain a unique parametrization of pairs of commuting
elements of $\isoc$ modulo $\isoc$ conjugation. For this purpose, let
$(R_\alpha,w_\alpha)$, $\alpha=1,2$, be commuting. Clearly then
$R_1R_2=R_2R_1$. From the $\su$ parametrization of $\soc$ given in
Refs.\cite{bargmann,carmeli}, it is straightforward to show that every matrix
$R\in \soc$ can be
written in the form $\exp(v^I K_I)$, where $v^I v_I \ge -\pi^2$ and $K_I$ are
the $3\times3$ matrices spanning the adjoint representation of the Lie algebra
of $\so$, ${\left(K_J\right)}^I{}_K=\epsilon^I{}_{JK}$. The only redundancy in
this parametrization of $\soc$ in terms of the Lorentz vector $v^I$ is that
when $v^I v_I = -\pi^2$, $v^I$ and $-v^I$ give the same element of $\soc$.
Writing now $R_\alpha=\exp(v_\alpha^I K_I)$ in this fashion, one sees that the
two Lorentz vectors $v_\alpha^I$ must be proportional to each other. One has
then four qualitatively different cases, depending on whether the two Lorentz
vectors are spacelike, timelike, null but nonvanishing, or vanishing.

{\it Case\/} (i). Suppose first that the vectors $v_\alpha^I$ are spacelike,
or that one of them is spacelike and the other vanishes. This means that
$R_\alpha$ are boosts with at least one having a nonvanishing boost
parameter. The holonomies $(R_\alpha,w_\alpha)$ can be conjugated to the
form
\begin{equation}
R_\alpha =
\exp(\lambda_\alpha K_2) =
\left(
\begin{array}{ccc}
\cosh \lambda_\alpha & \sinh \lambda_\alpha & 0 \\
\sinh \lambda_\alpha & \cosh \lambda_\alpha & 0 \\
0 & 0 & 1
\end{array}
\right)
\ \ , \ \
w_\alpha =
\left(
\begin{array}{c}
0 \\
0 \\
a_\alpha
\end{array}
\right)
\ \ ,
\label{i}
\end{equation}
where the four parameters $\lambda_\alpha$ and $a_\alpha$ can take
arbitrary values except that $\lambda_\alpha$ are by assumption not both
equal to zero. The only remaining redundancy in this parametrization is that
conjugation by the rotation $R=\exp(\pi K_0)$ reverses the signs of all the
four parameters. Thus, a unique parametrization is obtained from (\ref{i})
through the identification
$(\lambda_\alpha,a_\alpha)\sim(-\lambda_\alpha,-a_\alpha)$.  We denote this
part of $\miso$ by~$\ms$.

{\it Case\/} (ii). Suppose then that the vectors $v_\alpha^I$ are timelike,
or that one of them is timelike and the other vanishes. This means that
$R_\alpha$ are rotations, with at least one differing
from the identity rotation. The holonomies can be conjugated to the form
\begin{equation}
R_\alpha =
\exp({\tilde\lambda}_\alpha K_0) =
\left(
\begin{array}{ccc}
1 & 0 & 0 \\
0 & \cos {\tilde\lambda}_\alpha & -\sin {\tilde\lambda}_\alpha \\
0 & \sin {\tilde\lambda}_\alpha & \cos {\tilde\lambda}_\alpha
\end{array}
\right)
\ \ , \ \
w_\alpha =
\left(
\begin{array}{c}
- {\tilde a}_\alpha \\
0 \\
0
\end{array}
\right)
\ \ ,
\label{ii}
\end{equation}
where ${\tilde\lambda}_\alpha$ and ${\tilde a}_\alpha$ are four parameters,
with $|{\tilde\lambda}_\alpha|\le\pi$ and ${\tilde\lambda}_\alpha$ not
both equal to zero. This parametrization becomes unique when we
interpret ${\tilde\lambda}_\alpha$ as angular parameters, identified
according to
$({\tilde\lambda}_1,{\tilde\lambda}_2) \sim
({\tilde\lambda}_1 + 2\pi m,{\tilde\lambda}_2 + 2\pi n)$
for $m,n$ in~${\bf Z}$.
We denote this part of $\miso$ by~$\mt$.

{\it Case\/} (iii). Suppose then that the vectors $v_\alpha^I$ are null,
with at least one of them nonvanishing. This means that $R_\alpha$ are
null rotations, with at least one differing from the identity
rotation. The holonomies can now be conjugated to the form
\begin{equation}
\begin{array}{rcl}
R_1 &=&
\exp \left( \cos\theta (K_0+K_2) \right) =
\left(
\begin{array}{ccc}
1 + \half \cos^2\theta & \cos\theta & - \half \cos^2\theta \\
\cos\theta & 1 & - \cos\theta \\
\half \cos^2\theta & \cos\theta & 1 - \half \cos^2\theta
\end{array}
\right)
\\
\noalign{\bigskip}
w_1 &=&
\casehalf p_\theta \cos\theta
\left(
\begin{array}{c}
1 + \onethird \cos^2\theta \\
\cos\theta \\
-1 + \onethird \cos^2\theta
\end{array}
\right)
-
\casehalf p_r \sin\theta
\left(
\begin{array}{c}
1 \\
0 \\
1
\end{array}
\right)
\\
\noalign{\bigskip}
R_2 &=&
\exp \left( \sin\theta (K_0+K_2) \right) =
\left(
\begin{array}{ccc}
1 + \half \sin^2\theta & \sin\theta & - \half \sin^2\theta \\
\sin\theta & 1 & - \sin\theta \\
\half \sin^2\theta & \sin\theta & 1 - \half \sin^2\theta
\end{array}
\right)
\\
\noalign{\bigskip}
w_2 &=&
\casehalf p_\theta \sin\theta
\left(
\begin{array}{c}
1 + \onethird \sin^2\theta \\
\sin\theta \\
-1 + \onethird \sin^2\theta
\end{array}
\right)
+
\casehalf p_r \cos\theta
\left(
\begin{array}{c}
1 \\
0 \\
1
\end{array}
\right)
\ \ .
\end{array}
\label{iii}
\end{equation}
Here $\theta$ is identified with period~$2\pi$, and $p_\theta$ and $p_r$
are arbitrary. It can be verified that this parametrization is unique; the
reason for the nomenclature will become clear shortly. We denote this part
of $\miso$ by~$\mn$.

{\it Case\/} (iv). The remaining case is when $v_\alpha^I$ both vanish,
so that $R_\alpha$ are identity Lorentz transformations and
$(R_\alpha,w_\alpha)$ are purely translational. We denote this part of
$\miso$ by~$\mo$. There are several subcases depending on the
spacelike/timelike/null character of $w_\alpha$ and the plane
or line which they span. For example, when $w_\alpha$ span a spacelike
plane, the holonomies can be uniquely conjugated to the form
\begin{equation}
w_1 =
\left(
\begin{array}{c}
0 \\
r_1 \\
0
\end{array}
\right)
\ \ , \ \
w_2 =
\left(
\begin{array}{c}
0 \\
r_2 \cos\phi \\
r_2 \sin\phi
\end{array}
\right)
\ \ ,
\label{iv}
\end{equation}
where $r_\alpha>0$ and $0<\phi<\pi$. The global structure of $\mo$ is
fairly complicated, and we shall not need an explicit parametrization
for all the different subcases.

We have thus shown that $\miso$ is the disjoint union of $\ms$, $\mt$,
$\mn$, and~$\mo$. The topology on $\miso$ is induced from that of the
manifold $\isoc\times\isoc \simeq T^2\times{\bf R}^{10}$, from which it
is clear that $\miso$ is connected. We shall now examine how close $\miso$
comes to being a four-dimensional manifold, and what structure $\miso$
inherits from the action~(\ref{3-action}).

The set $\ms$ is open in~$\miso$, and it is a four-dimensional manifold with
topology $S^1\times {\bf R}^3$. A pair $\left( {\bar
A}_a^I , {\bar e}_{aI} \right)$ with the holonomies (\ref{i}) is given by
\begin{equation}
\begin{array}{rcl}
{\bar A}^2 &=& \lambda_\alpha dx^\alpha
\\
{\bar e}_2 &=& a_\alpha dx^\alpha
\ \ ,
\end{array}
\label{pair_s}
\end{equation}
with all other components vanishing. Here, and from now on, $(x^1,x^2)$
will denote a pair of angular coordinates on~$\Sigma$, periodic with the
identifications
$(x^1,x^2) \sim (x^1+m,x^2+n)$ for $m,n$ in~${\bf Z}$;
the Greek index
in Eqs.~(\ref{pair_s}) and henceforth should therefore be
understood not as abstract index notation but as referring to this
particular coordinate system. From the discussion of the holonomies above
it is clear that the fields (\ref{pair_s}) define an almost unique gauge
fixing, the only degeneracy being that the transformation
$(\lambda_\alpha,a_\alpha)
\mapsto
(-\lambda_\alpha,-a_\alpha)$ gives a
gauge-equivalent pair. Pulling the action (\ref{2+1-action}) back
to these spatially homogeneous fields, the induced Poisson
brackets on the parameters are found to be
\begin{equation}
\{ \lambda_1, a_2 \} = - \{ \lambda_2, a_1 \} = 1
\ \ .
\end{equation}
$\ms$ has therefore the structure of a phase space, and it is naturally
viewed as the cotangent bundle over the punctured plane. The symplectic
form is
\begin{equation}
\Omega_s = - \epsilon^{\alpha\beta} d a_\alpha \wedge
d\lambda_\beta
\ \ ,
\label{symp_s}
\end{equation}
where $\epsilon^{\alpha\beta}$ is the antisymmetric symbol
with the convention $\epsilon^{12}=1$.

The set $\mt$ is open in $\miso$, and it is a four-dimensional
manifold with the topology of a punctured torus times~${\bf R}^2$. A
pair $\left( {\bar A}_a^I , {\bar e}_{aI} \right)$ with
the holonomies (\ref{ii}) is given by
\begin{equation}
\begin{array}{rcl}
{\bar A}^0 &=& {\tilde \lambda}_\alpha dx^\alpha
\\
{\bar e}_0 &=& {\tilde a}_\alpha dx^\alpha
\ \ .
\end{array}
\label{pair_t}
\end{equation}
It is clear that this pair defines an almost unique gauge fixing, the only
degeneracy being that adding multiples of $2\pi$ to ${\tilde
\lambda}_\alpha$ gives gauge equivalent pairs. It can be verified as above
that $\mt$ inherits the symplectic structure
\begin{equation}
\Omega_t = - \epsilon^{\alpha\beta} d {\tilde a}_\alpha \wedge
d{\tilde\lambda}_\beta
\ \ ,
\label{symp_t}
\end{equation}
and $\mt$ can therefore be naturally viewed as the cotangent bundle
over the punctured torus.

The set $\mn$ is a three-dimensional manifold with topology
$S^1\times {\bf R}^2$. It is not open in~$\miso$, and the neighborhoods
of every point in $\mn$ intersect both $\ms$ and~$\mt$. A pair $\left(
{\bar A}_a^I , {\bar e}_{aI} \right)$ with the holonomies (\ref{iii}) is
given by
\begin{equation}
\begin{array}{rcl}
{\bar A}^0 &=& {\bar A}^2 =
du
\\
{\bar e}_0 &=&
- \casehalf p_r
dv
- \casehalf p_\theta
du
\\
{\bar e}_2 &=&
\casehalf p_r
dv
- \casehalf p_\theta
du
\ \ ,
\end{array}
\label{pair_n}
\end{equation}
where we have introduced the notation
\begin{equation}
\begin{array}{rcl}
du &=&
\cos\theta \, dx^1 + \sin\theta \, dx^2
\\
dv &=&
-\sin\theta \, dx^1 + \cos\theta \, dx^2
\ \ .
\end{array}
\label{uv}
\end{equation}
This pair clearly defines a unique gauge fixing.

Finally, the set $\mo$ is neither open in $\miso$ nor a manifold. It
however contains several subsets that are three-dimensional manifolds,
such as the set (\ref{iv}) homeomorphic to~${\bf R}^3$. A pair $\left(
{\bar A}_a^I , {\bar e}_{aI} \right)$ with the appropriate holonomies is
obtained by setting ${\bar A}_a^I=0$ and taking ${\bar e}^I_j dx^j$ equal
to the components of the column vector valued one-form $w_\alpha
dx^\alpha$. Note that all neighborhoods of the point $R_\alpha=1$,
$w_\alpha=0$ intersect both $\ms$, $\mt$, and~$\mn$.

The emerging picture of $\miso$ is thus that of two four-dimensional
manifolds, $\ms$ and~$\mt$, glued together by the three-dimensional
manifold $\mn$ and by the set $\mo$ which is somewhere close to being a
three-dimensional manifold. $\miso$ itself is not a manifold. One
way to see this is to consider the set~$\bmt$ (the closure of~$\mt$).
It is easily seen that the holonomies in $\bmt$ are given by
Eqs.~(\ref{ii}) with arbitrary values of ${\tilde\lambda}_\alpha$
and~${\tilde a}_\alpha$. $\bmt$ is clearly a manifold. It has the
symplectic structure given by the expression~(\ref{symp_t}), and it can be
naturally viewed as the cotangent bundle over the torus. Suppose now that
$\miso$ were a manifold; if so, it would have to be four-dimensional.
Consider a point $x\in\bmt\cap\mo$. Let $U\simeq{\bf R}^4$ be a
neighborhood of $x$ in~$\miso$. Then $U\cap\bmt$ is open in~$\bmt$. As
$\bmt$ is a four-manifold, there exists a set $V\subset U\cap\bmt$ such
that $x\in V$ and $V\simeq{\bf R}^4$. Then, $V\subset U$ implies that $V$
is open as a subset of~$U$; hence $V$ is open in~$\miso$. This however  is
a contradiction, since any neighborhood of $x$ in $\miso$  contains points
that are not in~$\bmt$. Therefore $\miso$ cannot be a manifold.

Note incidentally that the set $\bms$ (closure of~$\ms$), where the
holonomies are given by Eqs.~(\ref{i}) with arbitrary values of
$\lambda_\alpha$ and~$a_\alpha$, is not a manifold. The reason is that the
parametrization of $\bms$ by Eqs.~(\ref{i}) is unique only after the
identification $(\lambda_\alpha,a_\alpha)\sim(-\lambda_\alpha,-a_\alpha)$,
which has a fixed point at $\lambda_\alpha=a_\alpha=0$. This makes
$\bms$ an orbifold\cite{orbifold}, or a conifold\cite{conifold}.

There exists, nevertheless, an open subset of $\miso$ that contains
both $\ms$ and $\mt$ and {\it is\/} a manifold: it is
$\miso\setminus\mo=\mt\cup\mn\cup\ms$.
To show this, let us first perform on $\ms$
the coordinate transformation
\begin{equation}
\begin{array}{rcl}
\lambda_1 &=& {(2 \sinh 2r)}^{1/2} \cos\theta
\\
\lambda_2 &=& {(2 \sinh 2r)}^{1/2} \sin\theta
\\
a_1 &=& \displaystyle{
- {p_r \, {(2 \sinh 2r)}^{1/2} \sin\theta \over 2 \cosh 2r}
-
{p_\theta \cos\theta \over {(2 \sinh 2r)}^{1/2}}
}
\\
\noalign{\smallskip}
a_2 &=& \displaystyle{
{p_r \, {(2 \sinh 2r)}^{1/2} \cos\theta \over 2 \cosh 2r}
-
{p_\theta \sin\theta \over {(2 \sinh 2r)}^{1/2}}
}
\ \ ,
\end{array}
\label{s_to_r}
\end{equation}
where $r>0$, $\theta$ is periodic with period~$\pi$, and $p_r$ and
$p_\theta$ take arbitrary values. The symplectic structure (\ref{symp_s})
is given by
\begin{equation}
\Omega_s = d p_r \wedge dr + d p_\theta \wedge d\theta
\ \ .
\label{omega_s}
\end{equation}
Next, we use the Gauss constraint (\ref{G-constraint}) to perform on the
pair (\ref{pair_s}) a gauge transformation which amounts to a Lorentz boost
with rapidity $\half\ln(\coth r)$ in the internal (02) plane. Finally,
we use the curvature constraint (\ref{F-constraint}) to perform a second
gauge transformation, taking only the $I=1$ component of the gauge
transformation parameter to be nonvanishing and equal to
$p_\theta/(2\sinh 2r)$. The new pair is
\begin{equation}
\begin{array}{rcl}
{\bar A}^0 &=&
e^{-r} du
\\
{\bar A}^2 &=&
e^r du
\\
{\bar e}_0 &=&
\displaystyle{
-
{p_r e^{-r} dv  \over 2 \cosh 2r}
- {p_\theta du \over 2 \cosh r}
}
\\
\noalign{\medskip}
{\bar e}_2 &=&
\displaystyle{
{p_r e^r dv \over 2 \cosh 2r}
- {p_\theta du \over 2 \cosh r}
}
\ \ ,
\end{array}
\label{pair_trans}
\end{equation}
where $du$ and $dv$ are as in~(\ref{uv}). This pair is well-defined for
$-\infty<r<\infty$. For $r=0$ it reduces to~(\ref{pair_n}), provided
that $\theta$ is  regarded as periodic with period~$2\pi$. For $r<0$,
the pair (\ref{pair_trans}) is gauge-equivalent to the pair~(\ref{pair_t}),
via the coordinate transformation
\begin{equation}
\begin{array}{rcl}
{\tilde\lambda}_1 &=& {(-2 \sinh 2r)}^{1/2} \cos\theta
\\
{\tilde\lambda}_2 &=& {(-2 \sinh 2r)}^{1/2} \sin\theta
\\
{\tilde a}_1 &=& \displaystyle{
{{p_r \, (-2 \sinh 2r)}^{1/2} \sin\theta \over 2 \cosh 2r}
-
{p_\theta \cos\theta \over {(-2 \sinh 2r)}^{1/2}}
}
\\
\noalign{\smallskip}
{\tilde a}_2 &=& \displaystyle{
- {{p_r \, (-2 \sinh 2r)}^{1/2} \cos\theta \over 2 \cosh 2r}
-
{p_\theta \sin\theta \over {(-2 \sinh 2r)}^{1/2}}
}
\ \ ,
\end{array}
\label{r_to_t}
\end{equation}
again provided that $\theta$ is understood periodic with period~$2\pi$.
The required gauge transformation is the inverse of the one used when going
from (\ref{pair_s}) to~(\ref{pair_trans}). The periodic identifications of
${\tilde\lambda}_\alpha$ induce a set of corresponding identifications for
$(r,\theta,p_r,p_\theta)$  in the domain $r<0$ via~(\ref{r_to_t});
however, from the first two lines in (\ref{r_to_t}) it is seen that
no quadruplet $(r,\theta,p_r,p_\theta)$ in the range
$-\casehalf \mathop{\rm arsinh}\nolimits (\pi^2/2) <r<0$
is identified with another quadruplet in the same range.

Thus, the coordinates $(r,\theta,p_r,p_\theta)$ extend from $\ms$ through
$\mn$ to~$\mt$, with $\theta$ being periodic with period $\pi$ for $r>0$
and with period $2\pi$ for $r\leq0$. This shows that $\miso\setminus\mo$
is a manifold, although not a Hausdorff one, since points with $r=0$ and
$\theta$ differing by $\pi$ do not have disjoint neighborhoods.
$\miso\setminus\mo$ can therefore be viewed as the cotangent bundle whose
base space consists of the base space of~$\ms$ (punctured plane) and the
base space of~$\mt$ (punctured torus) glued together at the punctures; the
circle which provides the glue is the image of $\mn$ under the projection
that maps  the holonomies (\ref{iii}) to pure Lorentz transformations.
The circle joins to the base space of $\mt$ in a one-to-one fashion,
but the joining of the base space of $\ms$ to the circle is
two-to-one.

\subsection{$G=\isu$}
\label{subsec:su11}

We now turn to the case where ${\bar A}_a^I$ is an $\su$ connection. The
group $G$ is now $\isu$, and we denote the space of classical solutions
by~$\nisu$.

Recall that $\isu$ can be defined as the group of pairs $(U,H)$, where $U$
is an $\su$ matrix and $H$ is a Hermitian $2\times2$ matrix whose
diagonal elements are equal. The group multiplication law is
\begin{equation}
(U_2,H_2) \cdot (U_1,H_1) =
(U_2 U_1, U_2 H_1 U^\dagger_2 + H_2)
\ \ .
\end{equation}
Since $\su$ is a double cover of $\soc$, it follows that $\isu$ is a double
cover of $\isoc$\cite{bargmann,carmeli}.
Representing points in $\mink$ by the matrices
\begin{equation}
V =
\left(
\begin{array}{cc}
-T & Y + iX \\
 Y - iX & -T
\end{array}
\right)
\ \ ,
\end{equation}
where
$T$, $X$, and $Y$
are the usual Minkowski coordinates, the $\isoc$
action of $\isu$ on $\mink$ is given by
\begin{equation}
(U,H): \ \ V \mapsto U V U^\dagger + H
\ \ .
\end{equation}

We now wish to obtain a unique parametrization of pairs of commuting
elements of $\isu$ modulo $\isu$ conjugation. The main difference from the
previous subsection arises from the fact that not every element of
$\su$ can be written as the exponential of an element in the Lie
algebra.
Recall\cite{bargmann}
that a basis for the Lie algebra of $\su$ is given by
the three matrices
\begin{equation}
\tau_0 = - \casehalf i \sigma_3
\ \ , \ \
\tau_1 = \casehalf \sigma_1
\ \ , \ \
\tau_2 = \casehalf \sigma_2
\ \ ,
\end{equation}
where $\sigma_1$, $\sigma_2$ and $\sigma_3$ are the Pauli matrices. An
isomorphism with the Lie algebra of $\so$ is given by $\tau_I\mapsto
K_I$, where the  matrices $K_I$ were introduced in the previous subsection.
{}From the parametrization of $\su$ given in Refs.\cite{bargmann,carmeli}, it
is easily verified that every element of $\su$ can be written either
as $\exp(v^I\tau_I)$ or $-\exp(v^I\tau_I)$. For an $\su$ element for which the
corresponding Lorentz transformation is a rotation, the former expression is
sufficient (and still redundant); for an $\su$ element for which the
corresponding Lorentz transformation is a boost or a null rotation, there is a
unique parametrization by exactly one of the two expressions. Using this, the
analysis proceeds in analogy with that in the previous subsection. There are
again four qualitatively different cases.

{\it Case\/} (i). When the Lorentz transformations corresponding to the
$\su$ matrices are boosts, with at least one having a nonvanishing
boost parameter, the holonomies $(U_\alpha,H_\alpha)$ can be conjugated to
the form
\begin{equation}
U_\alpha =
{(-1)}^{\eta_\alpha}
\exp(\lambda_\alpha \tau_2) =
{(-1)}^{\eta_\alpha}
\left(
\begin{array}{cc}
\cosh (\lambda_\alpha/2) & -i\sinh (\lambda_\alpha/2) \\
i\sinh (\lambda_\alpha/2) & \cosh (\lambda_\alpha/2)
\end{array}
\right)
\ \ , \ \
H_\alpha =
\left(
\begin{array}{cc}
0 & a_\alpha \\
a_\alpha & 0
\end{array}
\right)
\ \ ,
\label{ui}
\end{equation}
where the four parameters $\lambda_\alpha$ and $a_\alpha$ can take
arbitrary values except that $\lambda_\alpha$ are not both equal to zero,
and the parameters $\eta_\alpha$ take the discrete values $0$ and~$1$.
A unique parametrization is obtained from (\ref{ui}) through the
identification $(\lambda_\alpha,a_\alpha)\sim(-\lambda_\alpha,-a_\alpha)$.
We denote these four parts of $\nisu$ respectively by~$\ns$, where the
upper indices correspond to the four possible combinations of
$\eta_\alpha$ in Eq.~(\ref{ui}).

{\it Case\/} (ii). When the Lorentz transformations corresponding to the
$\su$ matrices are rotations, with at least one differing from the
identity rotation, the holonomies $(U_\alpha,H_\alpha)$ can be conjugated
to the form
\begin{equation}
U_\alpha =
\exp({\tilde\lambda}_\alpha \tau_0) =
\left(
\begin{array}{cc}
\exp ( -i {\tilde\lambda}_\alpha/2) & 0 \\
0 & \exp ( i {\tilde\lambda}_\alpha/2)
\end{array}
\right)
\ \ , \ \
H_\alpha =
\left(
\begin{array}{cc}
{\tilde a}_\alpha & 0 \\
0 & {\tilde a}_\alpha
\end{array}
\right)
\ \ ,
\label{uii}
\end{equation}
where ${\tilde\lambda}_\alpha$ and ${\tilde a}_\alpha$ are four parameters,
arbitrary except that $({\tilde\lambda}_1,{\tilde\lambda}_2)\neq(2\pi
m,2\pi n)$ for $m,n$ in~${\bf Z}$. This parametrization becomes unique
when the parameters ${\tilde\lambda}_\alpha$ are identified according to
$({\tilde\lambda}_1,{\tilde\lambda}_2) \sim  ({\tilde\lambda}_1 +
4\pi m,{\tilde\lambda}_2 + 4\pi n)$ for $m,n$ in~${\bf Z}$.
We denote this part of $\nisu$ by~$\nt$.

{\it Case\/} (iii). When the Lorentz transformations corresponding to the
$\su$ matrices are null rotations, with at least one differing from
the identity rotation, the holonomies $(U_\alpha,H_\alpha)$ can be
conjugated to the form
\begin{equation}
\begin{array}{rcl}
U_1 &=&
{(-1)}^{\eta_1}
\exp \left( \cos\theta (\tau_0+\tau_2) \right) =
{(-1)}^{\eta_1}
\left(
\begin{array}{cc}
1 - \half i \cos\theta & -\half i \cos\theta \\
\half i \cos\theta & 1 + \half i \cos\theta
\end{array}
\right)
\\
\noalign{\bigskip}
H_1 &=&
\casehalf p_\theta \cos\theta
\left(
\begin{array}{cc}
-1 - \onethird \cos^2\theta
& -1 + \onethird \cos^2\theta + i \cos\theta    \\
-1 + \onethird \cos^2\theta - i \cos\theta
& -1 - \onethird \cos^2\theta
\end{array}
\right)
-
\casehalf p_r \sin\theta
\left(
\begin{array}{cc}
-1 & \phantom{-}1 \\
\phantom{-}1 & -1
\end{array}
\right)
\\
\noalign{\bigskip}
U_2 &=&
{(-1)}^{\eta_2}
\exp \left( \sin\theta (\tau_0+\tau_2) \right) =
{(-1)}^{\eta_2}
\left(
\begin{array}{cc}
1 - \half i \sin\theta & -\half i \sin\theta \\
\half i \sin\theta & 1 + \half i \sin\theta
\end{array}
\right)
\\
\noalign{\bigskip}
H_2 &=&
\casehalf p_\theta \sin\theta
\left(
\begin{array}{cc}
-1 - \onethird \sin^2\theta
& -1 + \onethird \sin^2\theta + i \sin\theta    \\
-1 + \onethird \sin^2\theta - i \sin\theta
& -1 - \onethird \sin^2\theta
\end{array}
\right)
+
\casehalf p_r \cos\theta
\left(
\begin{array}{cc}
-1 & \phantom{-}1 \\
\phantom{-}1 & -1
\end{array}
\right)
\ \ .
\end{array}
\label{uiii}
\end{equation}
Here $\theta$ is identified with period~$2\pi$, and $p_\theta$ and $p_r$
are arbitrary. The discrete parameters $\eta_\alpha$ take the values
$0$ and~$1$ as in case~(i). It can be verified that this
parametrization is unique. We denote these four parts of $\nisu$ by~$\nn$.

{\it Case\/} (iv). The remaining case is when
$U_\alpha={(-1)}^{\eta_\alpha}$ and $H_\alpha$ are arbitrary.
We denote these four parts of $\nisu$ by~$\no$. As in the previous
subsection, there are several subcases depending on the
spacelike/timelike/null character of the two Lorentz-vectors encoded in
$H_\alpha$ and the plane or line which they span.

We therefore see that $\nisu$ is the disjoint union of the thirteen sets
$\nt$, $\ns$, $\nn$, and~$\no$. The topology on $\nisu$ is induced from
that of the manifold $\isu\times\isu \simeq T^2\times{\bf R}^{10}$. It is
clear that $\nisu$ is connected. We shall now examine the structure of
$\nisu$ in analogy with the analysis in the previous subsection.

The four sets $\ns$ are each open in~$\nisu$, and each is a
four-dimensional manifold with topology $S^1\times {\bf R}^3$. A pair
$\left( {\bar A}_a^I , {\bar e}_{aI} \right)$ with the holonomies
(\ref{ui}) is given by
\begin{equation}
\begin{array}{rcl}
{\bar A}^0 &=& -2\pi \eta_\alpha dx^\alpha
\\
{\bar A}^1 &=& - \sin(2\pi \eta_\beta x^\beta)
\lambda_\alpha dx^\alpha
\\
{\bar A}^2 &=&  \cos(2\pi \eta_\beta x^\beta)
\lambda_\alpha dx^\alpha
\\
{\bar e}_1 &=& - \sin(2\pi \eta_\beta x^\beta)
a_\alpha dx^\alpha
\\
{\bar e}_2 &=& \cos(2\pi \eta_\beta x^\beta)
a_\alpha dx^\alpha
\ \ ,
\end{array}
\label{upair_s}
\end{equation}
as can be verified by direct computation.\footnote{That the fields
(\ref{upair_s}) are homogeneous only in ${\cal N}_s^{00}$ is not
coincidental. A direct exponentiation shows that any flat $\isu$
connection with a
homogeneous global
cross section will have its holonomies
either in ${\cal N}_s^{00}$, ${\cal N}_n^{00}$, ${\cal N}_0^{00}$,
or~$\nt$.} The fields (\ref{upair_s}) clearly define an almost unique gauge
fixing, the only degeneracy being that the transformation
$(\lambda_\alpha,a_\alpha)
\mapsto
(-\lambda_\alpha,-a_\alpha)$ gives a gauge-equivalent pair. The induced
symplectic structure on each $\ns$ is given by  Eq.~(\ref{symp_s}). All the
spaces $\ns$ are therefore isomorphic to each other and to $\ms$ as
symplectic manifolds.

The set $\nt$ is open in $\nisu$, and it is a four-dimensional
manifold with the topology of a four times punctured torus times~${\bf
R}^2$. A pair $\left( {\bar A}_a^I , {\bar e}_{aI} \right)$ with the
holonomies (\ref{uii}) is given by Eqs.~(\ref{pair_t}). This pair clearly
defines an almost unique gauge fixing, the only  degeneracy being that
adding multiples of $4\pi$ to ${\tilde \lambda}_\alpha$ gives gauge
equivalent pairs. The symplectic structure on $\nt$ is given
by~(\ref{symp_t}). As a symplectic manifold, $\nt$ is therefore a four-fold
cover of~$\ms$. It can be naturally viewed as the cotangent bundle over the
four times punctured torus.

The sets $\nn$ are three-dimensional manifolds with topology
$S^1\times {\bf R}^2$. They are not open in $\nisu$, and the neighborhoods
of every point in $\nn$ intersect both $\ns$ (with the same values
of the indices) and~$\nt$. A pair $\left( {\bar A}_a^I ,
{\bar e}_{aI} \right)$ with the holonomies (\ref{uiii}) is given by
\begin{equation}
\begin{array}{rcl}
{\bar A}^0 &=&
du
- 2\pi (\eta_\alpha dx^\alpha)
\\
{\bar A}^1 &=&
- \sin(2\pi \eta_\beta x^\beta)
du
\\
{\bar A}^2 &=&
\cos(2\pi \eta_\beta x^\beta)
du
\\
{\bar e}_0 &=&
- \casehalf p_r
dv
- \casehalf p_\theta
du
\\
{\bar e}_1 &=&
- \casehalf \sin(2\pi \eta_\beta x^\beta)
\left(
p_r dv - p_\theta du
\right)
\\
{\bar e}_2 &=&
\casehalf \cos(2\pi \eta_\beta x^\beta)
\left(
p_r dv - p_\theta du
\right)
\ \ .
\end{array}
\label{upair_n}
\end{equation}
This pair clearly defines a unique gauge fixing in each of the~$\nn$.

Finally, the sets $\no$ are neither open in $\nisu$ nor manifolds.
A pair $\left( {\bar A}_a^I , {\bar e}_{aI} \right)$ with the appropriate
holonomies is obtained by setting ${\bar A}^0=- 2\pi \eta_\alpha
dx^\alpha$ and taking ${\bar e}^I_j$ to be given by
\begin{equation}
\exp(2\pi \eta_\beta x^\beta \tau_0) H_\alpha
\exp(-2\pi \eta_\beta x^\beta \tau_0) dx^\alpha
=
\left(
\begin{array}{cc}
-{\bar e}^0_j & {\bar e}^2_j + i {\bar e}^1_j
\\
{\bar e}^2_j - i {\bar e}^1_j & -{\bar e}^0_j
\end{array}
\right)
dx^j
\ \ .
\end{equation}

Thus, $\nisu$ consists of the four-manifolds $\ns$ respectively glued to
the four-manifold $\nt$ by the three-manifolds $\nn$ and the sets~$\no$.
{}From the previous subsection it is clear that the gluing of ${\cal
N}_s^{00}$ to $\nt$ by  ${\cal N}_n^{00}$ and ${\cal N}_0^{00}$ is
identical to the gluing of $\ms$ to $\mt$ by $\mn$ and~$\mo$, and it is
straightforward to verify that the same holds also for the three other
gluings. We can therefore view $\nisu$ as a four-fold covering of~$\miso$.
This means in particular that the open subset  $\nisu\setminus \left(
\bigcup_{\eta_1 \eta_2} \no \right) = \nt  \cup \left( \bigcup_{\eta_1
\eta_2} \nn \right)  \cup \left( \bigcup_{\eta_1 \eta_2} \ns \right)$
is a manifold that contains ${\cal N}_t$ and all four
${\cal N}_s^{\eta_1 \eta_2}$. This non-Hausdorff symplectic manifold is a
four-fold cover of $\miso\setminus\mo$, and it is naturally viewed as the
cotangent bundle over a non-Hausdorff base space.

\section{Spacetime metrics}
\label{sec:metrics}

In this section we shall briefly discuss the relation of spacetime metrics
to the classical solutions in Witten's connection formulation. We shall
first concentrate on the case where ${\bar A}_a^I$ is an $\so$ connection.
The case when ${\bar A}_a^I$ is an $\su$ connection will be discussed at
the end of the section.

In the metric formulation of 2+1 gravity one starts with the assumption that
the spacetime metric is nondegenerate. In the Hamiltonian metric
formulation on the manifold $\Sigma\times{\bf R}$, where $\Sigma$ is a
closed orientable two-manifold, one conventionally makes further the
assumption that the induced metric on $\Sigma$ is positive definite. As
mentioned in section~\ref{sec:action}, Witten's connection formulation does not
contain these assumptions, neither in its Lagrangian nor Hamiltonian
form. An immediate consequence of this and the spacetime diffeomorphism
invariance of the actions (\ref{3-action}) and (\ref{2+1-action}) is that local
statements about the degeneracy and signature of the induced metric on $\Sigma$
become gauge-dependent: when ${\bar e}_{aI}$ is not identically zero, one can
locally change the signature of the induced metric on $\Sigma$ by spacetime
diffeomorphisms that distort the constant $t$ surfaces sufficiently wildly. But
also local statements about the degeneracy of the {\it three-metric\/} in
Witten's theory are gauge dependent\cite{witten1}. The geometrical reason is
perhaps most easily understood from the observation that as the action
(\ref{3-action}) and the field equations (\ref{eom3d}) contain only (Lie
algebra valued) forms but no vectors, the action and the classical solutions
can be pulled back by more general maps from ${\bf R}\times\Sigma$ to itself
than just diffeomorphisms\cite{horowitz}. For example, consider a smooth map
from ${\bf R}\times\Sigma$ to itself that
takes an open neighborhood $U$ to a single point. Given any solution to
Witten's theory, the pull-back of this solution gives a solution for which
${\bar A}_a^I$ and ${\bar e}_{aI}$, and hence in particular the metric $g_{ab}=
{\bar e}_{aI} {\bar e}_{b}^I$, vanish in~$U$. If the map is homotopic to the
identity map, the two solutions related in this fashion are clearly gauge
equivalent. These considerations indicate that the relation of Witten's
connection formulation to spacetime metrics should be addressed in terms of
globally defined questions.

(Interestingly, in the Hamiltonian metric formulation it is possible,
and arguably even natural, to relax the condition that the spacetime metric
is nondegenerate while still keeping the metric on $\Sigma$ positive definite.
The geometrical reason for this is analogous\cite{teitelboim1}: the
lapse-function can be regarded as a one-form with respect to time
reparametrizations, and the
Hamiltonian action does not involve its inverse. Hence, the Hamiltonian
formulation allows one to slice nondegenerate spacetimes in a fashion which may
locally stay still or even go backwards in the proper time. For a discussion of
some implications of this, see Refs.\cite{teitelboim2,hall-hartle}.)

One well-defined question is to start from the spacetimes of the
conventional Hamiltonian metric formulation (that is, spacetimes with
nondegenerate three-metrics and positive definite induced metrics on
$\Sigma$) and ask to which solutions in the connection formulation they
correspond\cite{mess}. For $\Sigma=T^2$, these spacetimes fall into
two classes\cite{moncrief,mess}. In the first class the holonomies
are in $\mo$ and given by~(\ref{iv}). A global nondegenerate co-triad is
\begin{equation}
\begin{array}{rcl}
{\bar e}_0 &=& dt
\\
{\bar e}_1 &=& r_1 dx^1 + r_2 \cos\phi \, dx^2
\\
{\bar e}_2 &=& r_2 \sin\phi \, dx^2
\ \ ,
\end{array}
\label{static_triad}
\end{equation}
which is obtained from the degenerate co-triad mentioned in subsection
\ref{subsec:so21} via a gauge transformation generated by the
$F$-constraint. This spacetime is just $\mink$ divided by the two purely
translational $\isoc$ holonomies~(\ref{iv}). In the second class the
holonomies are in $\ms$ and given by (\ref{i}) with
$\epsilon^{\alpha\beta}a_\alpha\lambda_\beta\neq0$. A global nondegenerate
co-triad is
\begin{equation}
\begin{array}{rcl}
{\bar e}_0 &=& -dt
\\
{\bar e}_1 &=& t \lambda_\alpha dx^\alpha
\\
{\bar e}_2 &=& a_\alpha dx^\alpha
\ \ ,
\end{array}
\label{space_triad1}
\end{equation}
where $t>0$. This co-triad is obtained from the degenerate co-triad
(\ref{pair_s}) via a gauge transformation generated by the $F$-constraint.
The resulting spacetime can be constructed from the region $T>|X|$ of
$\mink$ by taking the quotient with respect to the two $\isoc$ holonomies
(\ref{i})\cite{carlip1}.

{}From these results one expects that more general nondegenerate
spacetime metrics corresponding to points in $\miso$ can be found in a
similar fashion, dividing $\mink$ or some subset of it by the two $\isoc$
holonomies. We shall now demonstrate that this is true for all points in
$\miso$ except a set of measure zero.
It appears plausible that the metrics
we shall exhibit are, up to diffeomorphisms, the most general nondegenerate
metrics on ${\bf R}\times T^2$ that correspond to the connection
formulation.

Let us first reconsider the points in $\ms$ for which
$\epsilon^{\alpha\beta}a_\alpha\lambda_\beta\neq0$. The holonomies
(\ref{i}) do not act properly discontinuously on all of~$\mink$, and
dividing $\mink$ by these holonomies gives a modest generalization of
Misner space\cite{haw-ell}. However, the holonomies do act freely and
properly discontinuously above (or below) the null plane $T=X$, and
similarly above (or below) the null plane $T=-X$. For concreteness,
consider the domain $T>-X$. It can be verified that the co-triad
\begin{equation}
\begin{array}{rcl}
{\bar e}_0 &=& -\casehalf dt + \casehalf (t-1) \lambda_\alpha dx^\alpha
\\
{\bar e}_1 &=& \casehalf dt - \casehalf (t+1) \lambda_\alpha dx^\alpha
\\
{\bar e}_2 &=& a_\alpha dx^\alpha
\ \ ,
\end{array}
\label{space_triad2}
\end{equation}
which is gauge-equivalent to that given in Eq.~(\ref{pair_s}),
gives the metric of the space obtained by taking the quotient of
the domain $T>-X$ with respect to the two holonomies~(\ref{i}). The
subdomains $T>|X|$ and $X>|T|$ are covered respectively by $t>0$ and $t<0$,
and the surface $T=X$ is at $t=0$. The $t=\hbox{constant}$ tori go from
timelike for $t<0$ via null at $t=0$ to spacelike for $t>0$. The situation
is  highly similar to that of Misner space\cite{haw-ell}. Note that the
induced metric on the $t=0$ torus is just the one given by the degenerate
co-triad~(\ref{pair_s}). The existence of co-triads corresponding to
(\ref{space_triad2}) has been previously pointed out by Unruh and
Newbury\cite{unruh-newbury}.

For~$\mo$, where the holonomies are purely translational, the situation is
straightforward. Whenever the two translations are linearly independent,
one easily finds a nondegenerate co-triad gauge equivalent to that
mentioned in subsection \ref{subsec:so21} such that the resulting
spacetime is simply $\mink$ divided by the two translations. The tori are
spacelike only in the case~(\ref{static_triad}).

Consider then~$\mn$. For $p_r\neq0\neq p_\theta$, a nondegenerate
co-triad gauge equivalent to
that given in Eq.~(\ref{pair_n}) is
\begin{equation}
\begin{array}{rcl}
{\bar e}_0 &=&
- \casehalf \left[
(p_r dv - 2t du) + p_\theta du \right]
\\
{\bar e}_1 &=&
dt
\\
{\bar e}_2 &=&
\casehalf \left[
(p_r dv - 2t du) - p_\theta du \right]
\ \ ,
\end{array}
\label{null_triad1}
\end{equation}
where $-\infty<t<\infty$. The
spacetime corresponding to (\ref{null_triad1})
is obtained by taking the quotient of
$\mink$ with respect to the holonomies~(\ref{iii})
(the action is free and properly discontinuous), and the
$t=\hbox{constant}$ tori are timelike.  For $p_r\neq0=p_\theta$ the
co-triad (\ref{null_triad1}) becomes degenerate, but a gauge equivalent
nondegenerate co-triad is
\begin{equation}
\begin{array}{rcl}
{\bar e}_0 &=&
- \casehalf
( p_r dv + dt )
\\
{\bar e}_1 &=&
- t du
\\
{\bar e}_2 &=&
\casehalf ( p_r dv - dt )
\ \ ,
\end{array}
\label{null_triad2}
\end{equation}
where $t>0$. The spacetime corresponding to (\ref{null_triad2})
is obtained by taking the quotient of the region $T>Y$ (or, equivalently,
the region $T<Y$) of $\mink$ with respect to the holonomies~(\ref{iii}).
The $t=\hbox{constant}$ tori are now null. The action of the holonomies is
free and properly discontinuous in this region but not on the plane $T=Y$.
The geometrical picture is perhaps most easily seen from the observation
that the holonomies (\ref{iii}) are respectively the exponential maps of
the vectors $\cos\theta \,\xi_1 + \sin\theta \,\xi_2$ and  $-\sin\theta
\,\xi_1 + \cos\theta \,\xi_2$, where $\xi_1$ and $\xi_2$ are the two
commuting Killing vectors on $\mink$ given by
\begin{equation}
\begin{array}{rcl}
\xi_1 &=&
(T-Y) (\partial/\partial X) + X
\Bigl[
(\partial/\partial T) + (\partial/\partial Y)
\Bigr]
+
\casehalf p_\theta
\Bigl[
(\partial/\partial T) - (\partial/\partial Y)
\Bigr]
\\
\noalign{\smallskip}
\xi_2 &=&
\casehalf p_r
\Bigl[
(\partial/\partial T) + (\partial/\partial Y)
\Bigr]
\ \ .
\end{array}
\end{equation}
Note that $\xi_1$ reduces to a generator of a pure null
rotation for $p_\theta=0$.

Consider finally~$\mt$. When
$\epsilon^{\alpha\beta}{\tilde a}_\alpha{\tilde\lambda}_\beta\neq0$, a
nondegenerate co-triad gauge equivalent to
that given in Eq.~(\ref{pair_t}) is
\begin{equation}
\begin{array}{rcl}
{\bar e}_0 &=& {\tilde a}_\alpha dx^\alpha
\\
{\bar e}_1 &=& -t {\tilde \lambda}_\alpha dx^\alpha
\\
{\bar e}_2 &=& dt
\ \ ,
\end{array}
\label{time_triad}
\end{equation}
where $t>0$. The spacetime corresponding to (\ref{time_triad}) is obtained
by removing the $T$ axis from~$\mink$, going to the universal covering
space, and taking the quotient with respect to the holonomies~(\ref{ii}).
The $t=\hbox{constant}$ tori are timelike.
This procedure of obtaining a spacetime metric from the pair (\ref{pair_t})
has, however, the feature that the metric depends on
the gauge choice made for the pair by more than just
spacetime diffeomorphisms. If, instead of the pair~(\ref{pair_t}),
one starts from a gauge equivalent pair in which multiples of $2\pi$ have
been added to~${\tilde\lambda}_\alpha$, one finds a
nondegenerate co-triad which is obtained from (\ref{time_triad}) by adding
the same multiples of $2\pi$ to~${\tilde\lambda}_\alpha$. For generic
values of the parameters, the metric obtained from this new
co-triad will not be diffeomorphic to that obtained
from~(\ref{time_triad}). This means that in $\mt$ the relation of the
connection description to spacetime metrics is considerably looser than
in $\mo$ and~$\ms$.

One attempt to make the relation between points in $\mt$ and spacetime
metrics more rigid might be to declare that the pair (\ref{pair_t})
should correspond to the spacetime obtained by removing the $T$ axis
from~$\mink$ and taking the quotient with respect to the
holonomies~(\ref{ii}), {\it without\/} going to the covering space. Such a
spacetime could be seen as  an attempt to characterize the equivalence
class  of the co-triads (\ref{time_triad}) under the gauge transformations
which add multiples of $2\pi$ to~${\tilde \lambda}_\alpha$.
Whereas this viewpoint appears self-consistent, it suffers from the
drawback that for generic values of the parameters the action of the
holonomies is not properly discontinuous, and the space obtained after
taking the quotient is nowhere near a manifold. This quotient
space would therefore not correspond to the notion of ``spacetime" as
usually understood.

To end this section we shall briefly indicate how the above picture is
modified when ${\bar A}_a^I$ is an $\su$ connection. Recall that $\nisu$
can be naturally understood as a four-fold cover of~$\miso$. Consider a
point $x \in \miso\setminus\mt$ that corresponds to nondegenerate
spacetime metrics as above. Let $y$ be a point in $\nisu\setminus\nt$
which is taken to $x$ by the covering map.  Then the nondegenerate
spacetime metrics obtained above from $x$ are also described by co-triads
corresponding to~$y$. For $y$ in $\no$, $\ns$, or $\nn$, appropriate
co-triads are obtained from those in Eqs.\
(\ref{static_triad})-(\ref{null_triad2}) by an internal rotation according
to ${\bar e}^I \mapsto
{\left[\exp(2\pi\eta_\beta x^\beta K_0)
\right]}^I
{\vphantom{\left[\exp(2\pi\eta_\beta x^\beta K_0)
\right]}}_J {\bar e}^J$.
In $\nt$ the situation differs from that in $\mt$ only in that the large
gauge transformations act on the co-triad (\ref{time_triad}) by adding to
${\tilde \lambda}_\alpha$ multiples of $4\pi$ rather than~$2\pi$. An
attempt to construct a spacetime characterizing such a gauge-equivalence
class of triads might now be  to remove the $T$ axis from~$\mink$, go to a
{\it double\/} cover, and take the quotient of this double cover with
respect to the transformations corresponding to the
holonomies~(\ref{uii}). Again, for generic values of the parameters this
quotient space is nowhere near a manifold.

\section{Quantum theories for $\Sigma=T^2$}
\label{sec:quantization}

We shall now explore formulations of quantum gravity
based on the solution spaces
$\miso$ and~$\nisu$.  Because these spaces
have neither a preferred set
of global coordinates
nor the structure of a cotangent
bundle over a manifold, this will
require some creativity.
We shall follow the technique of judiciously choosing the functions
that are to become operators in the quantum theory, in combination
with replacing $\miso$ and $\nisu$ by their simpler subsets. Some of
the resulting theories coincide with those presented in
Refs.\cite{five-a,AAbook2,carlip1,marolf}, whereas others are new.

Let us first recall the geometric quantization framework
of Refs.\cite{woodhouse,AAbook-geom} for quantizing the cotangent
bundle $T^*B$ of a manifold~$B$.
For any smooth functions
$f,g$ and vector fields $v,w$
on~$B$, one defines the commutation relations
\begin{equation}
\label{alg}
[f,g] = 0, \qquad [v,f] = -i \mathpounds_{v} f, \qquad [v,w]
= -i \{v,w\}_\mathpounds.
\end{equation}
Here, $\{v,w\}_\mathpounds$ stands for the Lie bracket of vector
fields and $\mathpounds_v$ is the Lie derivative along~$v$.
Together with the involution operation $\star$ defined by
complex conjugation, Eq.~(\ref{alg}) defines a
$\star$-algebra of functions and vector fields on~$B$.
One may now seek Hilbert space representations of
(\ref{alg}) which implement the $\star$-relations as Hermiticity
relations.

One representation acts on smooth square integrable
half-densities on~$B$.
Functions act by multiplication
and vector fields act by $-i$
times Lie differentiation.
Note that this representation is well-defined
even when $B$ is non-Hausdorff.
With respect to the inner product
$(\alpha, \beta) = \int_B {\bar \alpha} \beta$,
$\star$ is the adjoint operation when acting on
any function or any vector field,
with suitable fall-off conditions
imposed when $B$ is non-compact.
Completing the space of smooth half-densities in
the corresponding norm gives a representation
of the $\star$-algebra on $L^2(B)$.
Note that if $B$ is disconnected, this
representation is topologically reducible to a
direct sum of representations corresponding to the
connected components of~$B$. One can therefore
without loss of generality  assume $B$ to be
connected.

We would now like to apply this framework to some subsets of $\miso$
and $\nisu$ that were seen in section \ref{sec:solutions} to be cotangent
bundles. For~$\miso$, two natural subsets are $\mt$ and~$\ms$; as these
subsets are disconnected, considering them separately is equivalent
to considering the set $\mt\cup\ms$ which consists of all of $\miso$
except a set of measure zero. A third natural subset is
$\miso\setminus\mo$ which contains both $\ms$ and~$\mt$.
For~$\nisu$, the corresponding natural subsets
are~$\nt$, the four $\ns$, and $\nisu\setminus{\cal N}_0$, where ${\cal
N}_0 =\bigcup_{\eta_1 \eta_2} \no$.

When $T^*B$ is $\mt$ or $\ms$, or respectively
$\nt$ or $\ns$, the above quantization
is straightforward to implement. In particular, for
$\nt$ and ${\cal N}_s^{00}$ the resulting quantum
theories are equivalent to those given
in Refs.\cite{five-a,AAbook2}.
One further sees that replacing $\nt$ and $\mt$ by their closures
$\bnt$ and $\bmt$ does not change the resulting quantum
theories.\footnote{The difference between
$\nt$ and~$\bnt$ ($\mt$ and~$\bmt$, respectively)
becomes however more involved when one only promotes into
operators the elements of the classical $T$-algebra of
Refs.\cite{five-a,AAbook2}. See Ref.\cite{marolf}.}

For the rest of this section we shall concentrate on the cases where $T^*B
= \miso\setminus\mo$ or $T^*B = \nisu\setminus{\cal N}_0$. In these cases
the non-Hausdorff nature of $B$ leads to an unusual feature. Let us write
$B =  B_s \cup  B_n \cup B_t$, where $B_s$, $B_n$, and $B_t$ are the parts
of the base space that correspond to $\ms$, $\mn$, and $\mt$, or
respectively to ${\cal N}_s = \bigcup_{\eta_1\eta_2} \ns$,  ${\cal N}_n =
\bigcup_{\eta_1\eta_2} \nn$, and~$\nt$.  Then,
exponentials of
self-adjoint operators corresponding to complete vector fields which
are not tangent to $B_n$ are not defined on all densities.  In particular,
consider the map $\alpha:T^*B\to T^*B$, defined on $T^*B_s$
as the identity map, on $T^*B_t$ by
$({\tilde\lambda}_\alpha, {\tilde a}_\alpha) \mapsto
(-{\tilde\lambda}_\alpha, -{\tilde a}_\alpha)$,
and on the
fibers
over $B_n$ by
$(\theta,p_{\theta},p_r)\mapsto(\theta + \pi,p_{\theta},p_r)$,
where ${\tilde\lambda}_\alpha$,
${\tilde a}_\alpha$,
$\theta$, $p_r$, and $p_{\theta}$
were defined in section~\ref{sec:solutions}. Then, exponentiation of all
operators corresponding to complete vector fields that are invariant under
$\alpha$ is defined only on densities that are also invariant
under~$\alpha$.

If the usual type of quantum theory is to be
recovered, this problem must be corrected.  One
solution
is to promote to operators
only those functions and vector fields
that are invariant under~$\alpha$.  A
topologically irreducible representation of the
resulting algebra is given as above
when the operators are taken to act on half-densities
that are invariant under~$\alpha$.
Now, in the resulting quantum theory,
any complete smooth vector field may
be exponentiated to act on any half-density.
Since $B$ can be interpreted as a configuration
space, these vector fields
have an interpretation as momentum
operators. In particular, there
exist momentum operators whose
exponentials take half-densities with support
in $B_s$ to half-densities
whose support is in~$B_t$, and vice versa.
In this sense, then,
this quantum theory includes operators which induce
transitions between the theories of
Refs.\cite{five-a,AAbook2,carlip1}.
This is a result of the fact that
$B$ is a connected manifold.

Another approach is to consider the quotient of
$B$ and its cotangent bundle by
the map~$\alpha$.
Consider first the case $T^*B = \nisu\setminus{\cal N}_0$. The action of
$\alpha$ on $B$ is free and properly discontinuous, and the quotient
$(T^*B)/\alpha$ is a cotangent bundle over the manifold $B/\alpha$.
It is easily seen that $B/\alpha$ consists of a four times punctured
sphere glued to four punctured planes. The gluings at the
punctures are now one-to-one, and $B/\alpha$ is Hausdorff.
A Hilbert space representation of our
algebra on half-densities on $B/\alpha$ follows as
above, and the pull-back from $B/\alpha$
to $B$ provides an isomorphism with the representation
constructed above from functions,
vector fields, and half-densities on $B$ that are
invariant under~$\alpha$.

Consider then the case $T^*B = \miso\setminus\mo$. The action of
$\alpha$ on $B$ leaves fixed the three points
in $B_t$ for which
$({\tilde\lambda}_1,{\tilde\lambda}_2) =
(m\pi,n\pi)$, with $m,n$
in~${\bf Z}$ and not both even. Hence $B/\alpha$
is an orbifold with three singular points.
The quotient $(T^*B)/\alpha$
is not a bundle since the ``fibers" over the
singular points are not~${\bf R}^2$, but ${\bf R}^2$ with
opposite points identified.
Now let $B'$ be the complement in $B$ of the
three fixed points of~$\alpha$. Then $\alpha$ acts freely and properly
discontinuously on $B'$ and~$T^*B'$, and $B'/\alpha$ consists of a
punctured plane glued to a four times punctured sphere at one of the
punctures. The gluing at the puncture is one-to-one, so
$B'/\alpha$ is Hausdorff, and a representation of
our algebra on half-densities follows as above.  Note
that complete vector fields on $B'/\alpha$ pull back to
vector fields on $B$ that vanish rapidly near the three fixed points
of~$\alpha$. Since all smooth complete vector fields on
$B$ that are invariant under $\alpha$
also vanish rapidly at these three
points, a pull-back again provides
an isomorphism with the representation constructed from
functions, vector fields, and half-densities on $B$ that are
invariant under~$\alpha$.

In passing, we note that $\alpha$ is the map on $T^*B$
induced by inversion of the torus:
$x^{\beta} \mapsto -x^{\beta}$.
The above construction thus describes theories invariant
under this limited class of large diffeomorphisms.

Alternatively, we could return to
$T^*B = {\cal M}\setminus{\cal M}_0$ or
$T^*B = {\cal N}\setminus{\cal N}_0$ and
promote to operators only those
vector fields that are tangent to~$B_n$.
A Hilbert space representation of this algebra as above is
topologically reducible to a sum of two representations,
one on densities with support on $B_t$ and one with
support on~$B_s$.
(For $T^*B = \nisu\setminus{\cal N}_0$, the representation
with support on $B_s$ is further topologically reducible to
a sum of four representations corresponding to the four
disconnected components of~$B_s$.)
This again leads
to an analysis similar to that of
Refs.\cite{five-a,AAbook2,carlip1},
since the vector fields $T^1(\gamma)$ used in
Refs.\cite{five-a,AAbook2}
are tangent to~$B_n$.
However, because Refs.\cite{five-a,AAbook2}
promote to operators only
the functions $T^0(\gamma)$ and vector fields $T^1(\gamma)$ belonging to
the $T$-algebra, all of whom are invariant under~$\alpha$,
the representation obtained on $B_t$
is further reducible.

Taken to the extreme, selective promotion of functions to
operators becomes a new approach
by itself.
Let us consider
only those
smooth
functions on ${\cal M}$ or ${\cal N}$
that vanish on ${\cal M}_0$ or ${\cal N}_0$.
Examples are again
$T^0(\gamma)-1$
and $T^1(\gamma)$.
Such functions are essentially functions on
$T^*B = {\cal M}\setminus{\cal M}_0$
or $T^*B = {\cal N}\setminus{\cal N}_0$
and may again be interpreted as built from functions and
vector fields on~$B$.
The further restriction
to functions that vanish on $B_n$ and vector fields
tangent to $B_n$ again leads to the analysis of
Ref.\cite{AAbook2}.

\section{Conclusions and discussion}
\label{sec:discussion}

In this paper we have investigated Witten's connection formulation of
2+1 gravity, and its $\su$ modification, on the manifold ${\bf R}\times
T^2$. The space of classical solutions was shown to be connected, unlike
in the case where the torus is replaced by a higher genus surface. Also, we
saw that all points of this space except a set of measure zero correspond
to nondegenerate spacetime metrics on  ${\bf R}\times T^2$, even though not
all of these spacetimes admit slicings with spacelike tori.

Removing from the space of classical solutions a set of measure zero, we
obtained a manifold which is naturally viewed as the cotangent bundle
over a non-Hausdorff base space. This allowed us to explore various
possibilities for quantizing the theory. In addition to recovering the
quantum theories previously presented in
Refs.\cite{five-a,AAbook2,carlip1}, we exhibited a larger quantum theory
which contains the theories of Refs.\cite{five-a,AAbook2,carlip1} as its
parts. In particular, the larger theory contains operators that induce
transitions between the smaller theories.

While all our quantum theories incorporated invariance under the
connected component of the diffeomorphism group of the torus, the only
disconnected component considered was the one containing inversions of the
torus. If one attempts to build connection quantum theories that are
invariant under all diffeomorphisms, including the disconnected ones, one
encounters the fact that the full diffeomorphism group does not act even
remotely properly discontinuously on our base spaces. For example,
consider the action of the large diffeomorphisms on~$\ms$, given
by\cite{carlip1}
\begin{equation}
\left(
\begin{array}{c}
\lambda_1 \\
\lambda_2 \\
a_1 \\
a_2
\end{array}
\right)
\longmapsto
\left(
\begin{array}{cc}
M & 0 \\
0 & M
\end{array}
\right)
\left(
\begin{array}{c}
\lambda_1 \\
\lambda_2 \\
a_1 \\
a_2
\end{array}
\right)
\ \ ,
\label{ms_action}
\end{equation}
where $M\in\sltwoz$. Although the action (\ref{ms_action}) is properly
discontinuous and free in the domains
$\epsilon^{\alpha\beta}a_\alpha\lambda_\beta>0$ and
$\epsilon^{\alpha\beta}a_\alpha\lambda_\beta<0$, it is not
properly discontinuous on the surface
$\epsilon^{\alpha\beta}a_\alpha\lambda_\beta=0$. In particular, the
projection of the action (\ref{ms_action}) to the base space
$a_{\alpha}=0$ is not properly discontinuous. This makes the formulation
of connection quantum theories invariant under the large diffeomorphisms a
highly subtle question\cite{carlip1,carlip2,carlip3}.\footnote{In our
understanding, a subtlety in the approach of Refs.\cite{carlip2,carlip3}
arises from the domain of integration in Eq.~(3.3) of Ref.\cite{carlip2}.
We thank Steve Carlip for discussions on this point.}

In contrast, in the conventional metric theory the large diffeomorphisms
do act properly discontinuously (although not freely) on the configuration
space, and a quantum theory invariant under the large diffeomorphisms is
readily constructed\cite{hosoya-nakao2,carlip1}. This difference between
the metric theory and the connection theory is related to the fact that
the metric theory of Ref.\cite{carlip1} contains classically only the
region $\epsilon^{\alpha\beta}a_\alpha\lambda_\beta>0$  (or, equivalently,
the region  $\epsilon^{\alpha\beta}a_\alpha\lambda_\beta<0$)
of~$\ms$.

It should be noted that these subtleties with the large diffeomorphisms
have no counterpart with the higher genus surfaces. There, the
geometrodynamically relevant connected component of the classical solution
space to Witten's theory is the cotangent bundle over the \teich\ space
of~$\Sigma$, and the action of the large diffeomorphisms on the \teich\
space is properly discontinuous, although not free\cite{witten1,goldman2}.

It is straightforward to generalize our results to modifications of
Witten's theory where the gauge group of the connection ${\bar A}_a^I$ is
taken to be the $n$-fold covering group of $\soc$, $n>2$. The space of
classical solutions can be understood simply as the $n\times n$ cover
of~$\miso$, containing $n\times n$ counterparts of $\ms$, $\mn$ and $\mo$
attached to an $n\times n$ cover of~$\mt$. Similarly, if the gauge group
is taken  to be the universal covering group $\tsoc$ of $\soc$, the space
of classical solutions will consist of infinitely many counterparts of
$\ms$, $\mn$ and $\mo$ attached to the universal covering space of~$\mt$.
The discussion of the classical spacetime metrics for these theories is an
obvious generalization of that given in section~\ref{sec:metrics}.
(Note, however, that for $\tsoc$ there are no gauge transformations in the
connection formulation that would  add multiples of $2\pi$ to the
parameters ${\tilde \lambda}_\alpha$ in the co-triad~(\ref{time_triad}).)
Also, quantization of these theories is an obvious generalization of that
discussed in section~\ref{sec:quantization}.

In conclusion, we have seen that Witten's formulation of 2+1 gravity on
${\bf R}\times T^2$ differs qualitatively from this theory on
manifolds where the torus is replaced by a higher genus surface, both at
the classical and quantum levels. While the torus case admits quantum
theories that are reasonably close to the geometrodynamically relevant
quantum theories of the higher genus cases, it also admits quantum
theories that have no apparent analogue for higher genus.
Incorporating invariance under large diffeomorphisms in a quantum
theory in the torus case remains, however, a subtle question.

{\it Note added.} After the completion of this work we became aware of
Ref.\cite{nicolai}, in which the moduli space of flat $\su$ connections on the
torus is described in the context of $d=3, N=2$ supergravity.
This moduli space is the subspace of our $\nisu$ for which the $\isu$
holonomies are in $\su$.

\acknowledgments
We would like to thank Abhay Ashtekar, Steve Carlip, Alan Rendall,
Bill Unruh, and Madhavan Varadarajan for helpful discussions. This work
was supported in part by the NSF grants PHY90-05790 and PHY90-16733, and
by research funds provided by Syracuse University.


\newpage

\begin{references}

\bibitem[*]{jorma}
Electronic address: louko@convex.csd.uwm.edu.
Present address:
Department of Physics,
University of
Wisconsin-Milwaukee,
P.O.\ Box 413,
Milwaukee, WI~53201, USA.

\bibitem[\dag]{don}
Electronic address: marolf@hbar.phys.psu.edu.
Present address:
Department of Physics,
The Pennsylvania State University,
University Park,
PA~16802, USA.

\bibitem{deser1}
Deser S,
Jackiw R
and
't~Hooft, G
1984
{\it Ann.\ Phys.\ }(N.Y) {\bf 152} 220

\bibitem{deser2}
Deser S
and
Jackiw R
1984
{\it Ann.\ Phys.\ }(N.Y) {\bf 153} 405

\bibitem{martinec}
Martinec E
1984
{\it Phys.\ Rev.\ \rm D \bf 30} 1198

\bibitem{witten1}
Witten E
1988
{\it Nucl.\ Phys.\  \bf B311} 46

\bibitem{bengtsson1}
Bengtsson I
1989
{\it Phys.\ Lett.\ \bf 220B} 51

\bibitem{five-a}
Ashtekar A, Husain V, Rovelli C, Samuel J and Smolin L
1989
{\it Class.\ Quantum Grav.\ \bf 6} L185

\bibitem{AAbook2}
Ashtekar A
1991
{\it Lectures on Non-Perturbative Canonical Gravity\/}
(World Scientific, Singapore) Chapter 17

\bibitem{hosoya-nakao2}
Hosoya A and Nakao K
1990
{\it Prog.\ Theor.\ Phys.\ \bf 84} 739

\bibitem{carlip1}
Carlip S
1990
{\it Phys.\ Rev.\ \rm D \bf 42} 2647

\bibitem{carlip2}
Carlip S
1992
{\it Phys.\ Rev.\ \rm D \bf 45} 3584
[Erratum (1993) {\it Phys.\ Rev.\ \rm D \bf 47} 1729]

\bibitem{carlip3}
Carlip S
1993
{\it Phys.\ Rev.\ \rm D \bf 47} 4520

\bibitem{mano}
Manojlovi\'c N and
Mikovi\'c A
1992
{\it  Nucl.\ Phys.\ \bf B385} 571

\bibitem{barbero}
Barbero J F and Varadarajan M
1993
``The Phase Space of 2+1 Dimensional Gravity in the Ashtekar Formulation",
Syracuse Preprint SU-GP-93/64, gr-qc/9307006

\bibitem{carlip-water}
Carlip S
1993
``Six ways to quantize (2+1) dimensional gravity,"
Preprint
UCD-93-15,
NSF-ITP-93-63,
gr-qc/9305020,
in {\it Proceedings of the Fifth Canadian
Conference on General Relativity and Relativistic Astrophysics\/}
(World Scientific, Singapore) (to be published)

\bibitem{goldman1}
Goldman W M
1984
{\it Adv.\ Math.\ \bf 54} 200

\bibitem{goldman2}
Goldman W M
1985
in {\it Geometry and Topology\/}
(Lecture Notes in Mathematics, Vol.~1167),
eds.\ Alexander J and Harer J
(Springer, Berlin)

\bibitem{moncrief}
Moncrief V
1989
{\it J. Math.\ Phys.\ \bf 30} 2907

\bibitem{mess}
Mess G
1990
``Lorentz spacetimes of constant curvature,"
Institutes des Hautes Etudes Scientifiques Preprint
IHES/M/90/28

\bibitem{hosoya-nakao1}
Hosoya A and Nakao K
1990
{\it Class.\ Quantum Grav.\ \bf 7} 163

\bibitem{marolf}
Marolf D M
1993
``Loop Representations for 2+1 Gravity on a Torus,"
{\it Class.\ Quantum Grav.\ }(in press)

\bibitem{anderson}
Anderson A
1993
{\it Phys.\ Rev.\ D \bf 47} 4458

\bibitem{rov-smolin}
Rovelli C and Smolin L
1990
{\it Nucl.\ Phys.\ \bf B331} 80

\bibitem{AAbook-loop}
Ref.\cite{AAbook2}
Chapter 16

\bibitem{romano}
Romano J D
1993
{\it Gen.\ Rel.\ Grav.\ \bf 25} 759

\bibitem{bargmann}
Bargmann V
1947
{\it Ann.\ Math.\ \bf 38} 568

\bibitem{carmeli}
Carmeli M
1977
{\it Group Theory and General Relativity\/}
(McGraw-Hill, New York)
Chapter 3

\bibitem{orbifold}
Dixon L, Harvey J, Vafa C and Witten E
1986
{\it Nucl.\ Phys.\  \bf B274} 285

\bibitem{conifold}
Schleich K and Witt D M
1993
{\it Nucl.\ Phys.\ \bf B402} 411

\bibitem{horowitz}
Horowitz G T
1991
{\it Class.\ Quantum Grav.\ \bf 8} 587

\bibitem{teitelboim1}
Teitelboim C
(1982)
{\it Phys.\ Rev.\ \rm D \bf 25} 3159

\bibitem{teitelboim2}
Teitelboim C
1982
in {\it Quantum structure of spacetime\/}
eds.\ Duff M J and Isham C J
(Cambridge University Press, Cambridge)

\bibitem{hall-hartle}
Halliwell J J and Hartle J B
1991
{\it Phys.\ Rev.\ \rm D \bf 43} 1170

\bibitem{haw-ell}
Hawking S W and Ellis G F R
1973
{\it The Large Scale
Structure of Space-time\/}
(Cambridge University Press, Cambridge)

\bibitem{unruh-newbury}
Unruh W G and Newbury P
1993
{\it Phys.\ Rev.\ \rm D \bf 48} 2686

\bibitem{woodhouse}
Woodhouse N M J
1980
{\it Geometric Quantization\/}
(Clarendon Press, Oxford)

\bibitem{AAbook-geom}
Ref.\cite{AAbook2}
Chapter 10 and Appendix C

\bibitem{nicolai}
de~Wit B, Matschull H J and Nicolai H
1993
``Physical States in $d=3$, $N=2$ Supergravity"
Preprint DESY 93-125, THU-93/19, gr-qc/9309006


\end{references}
\end{document}